\documentclass[12pt]{article}
\usepackage{amsmath,amsfonts,amssymb}

\begin{document}
\thispagestyle{empty}

\def\theequation{\arabic{section}.\arabic{equation}}
\def\a{\alpha}
\def\b{\beta}
\def\g{\gamma}
\def\d{\delta}
\def\dd{\rm d}
\def\e{\epsilon}
\def\ve{\varepsilon}
\def\z{\zeta}
\def\B{\mbox{\bf B}}\def\cp{\mathbb {CP}^3}

\newcommand{\h}{\hspace{0.5cm}}

\begin{titlepage}
\vspace*{1.cm}
\renewcommand{\thefootnote}{\fnsymbol{footnote}}
\begin{center}
{\Large \bf Some semiclassical structure constants for
$AdS_4\times CP^3$}
\end{center}
\vskip 1.2cm \centerline{\bf Changrim  Ahn and Plamen Bozhilov
\footnote{On leave from Institute for Nuclear Research and Nuclear
Energy, Bulgarian Academy of Sciences, Bulgaria.}}

\vskip 10mm

\centerline{\sl Department of Physics} \centerline{\sl Ewha Womans
University} \centerline{\sl DaeHyun 11-1, Seoul 120-750, S. Korea}
\vspace*{0.6cm} \centerline{\tt ahn@ewha.ac.kr,
bozhilov@inrne.bas.bg}

\vskip 20mm

\baselineskip 18pt

\begin{center}
{\bf Abstract}
\end{center}

We compute structure constants in three-point functions of three string states in
$AdS_4\times CP^3$
in the framework of the semiclassical approach.
We consider HHL correlation functions where two of the states are ``heavy" string states of {\it finite-size}
giant magnons carrying one or two angular momenta and the other one corresponds to such ``light"
states as dilaton operators with {\it non-zero} momentum,
primary scalar operators, and singlet scalar operators with higher string levels.

\end{titlepage}
\newpage
\baselineskip 18pt

\def\nn{\nonumber}
\def\tr{{\rm tr}\,}
\def\p{\partial}
\newcommand{\non}{\nonumber}
\newcommand{\bea}{\begin{eqnarray}}
\newcommand{\eea}{\end{eqnarray}}
\newcommand{\bde}{{\bf e}}
\renewcommand{\thefootnote}{\fnsymbol{footnote}}
\newcommand{\be}{\begin{eqnarray}}
\newcommand{\ee}{\end{eqnarray}}

\vskip 0cm

\renewcommand{\thefootnote}{\arabic{footnote}}
\setcounter{footnote}{0}

\setcounter{equation}{0}
\section{Introduction}

The AdS/CFT duality between string/M theories on Anti-de Sitter (AdS) background
and conformal field theories (CFTs) on its boundary has been
a most productive research direction ever since it was proposed \cite{AdS/CFT}.
As a strong-weak coupling duality, integrability has played crucial roles in
non-perturbative computations \cite{review}.
It has been applied to compute conformal dimensions of the
CFTs and energies of corresponding string states.
A natural next challenge is to utilize integrability to compute structure constants
which determine three-point functions.

A promising recent progress is so-called hexagon amplitude approach \cite{BKV}.
The structure constants are given by sums of hexagon amplitudes, which can be determined exactly
in all orders of 't Hooft coupling constant $\lambda$.
This approach has proved effective in the weak coupling limit \cite{weakhexagon}.
However, technical difficulties such as summing up all intermediate states, finite-size effects, etc.
become substantial in the strong coupling limit \cite{stronghexagon}.

For two-heavy and one-light operators in the semiclassical limit $\lambda\gg 1$,
the ``HHL'' three-point functions can be obtained from explicit evaluation of light vertex operator
with the heavy string configurations \cite{hhl,RT2010}.
In spite of limited applicability, this method is useful to obtain structure constants when the
heavy operators have large but finite $J\gg\sqrt{\lambda}$ values.
The resulting HHL functions show exponential corrections $e^{-J/\sqrt{\lambda}}$,
which can be related to exact $S$-matrix, hence the integrability \cite{BajJan}.

Type IIA string theory on $AdS_4 \times CP^3$ background is dual
to $\mathcal{N}=6$ super Chern-Simons theory in three space-time dimensions, known as ABJM theory \cite{ABJM}.
Classical integrability \cite{claint} and
giant magnon solutions have been studied in \cite{GHO}-\cite{AA0811}.
The HHL 3-point functions have been computed for various string states
in $AdS_4 \times CP^3$ \cite{AR2010}-\cite{A2013}.
In this paper, we will focus on finite-size effects of some normalized structure constants in
$AdS_4\times CP^3$ in semiclassical limit where the heavy string
states are {\it finite-size} giant magnons.
We also consider various different light string states,
such as dilaton operators with {\it non-zero} momentum,
primary scalar operators, and singlet scalar operators on higher string levels.

The paper is organized as follows. In sect.2, we introduce
preliminary contents along with various giant magnons on $CP^3$.
We present the HHL functions of two giant magnons with
dilaton operator with non-zero momentum in sect.3, with 
primary scalar operators in sect.4, and with singlet
scalar operators on higher string levels in sect.5. We
conclude the paper in sect.6.

\setcounter{equation}{0}
\section{Preliminaries}
\subsection{Structure Constants}
It is known that correlation functions of CFTs can be determined  in principle in terms of the basic
conformal data $\{\Delta_i,C_{ijk}\}$, where $\Delta_i$ are the
conformal dimensions defined by two-point normalized correlation functions
\begin{equation}\nn
\left\langle{\cal O}^{\dagger}_i(x_1){\cal O}_j(x_2)\right\rangle=
\frac{\delta_{ij}}{|x_1-x_2|^{2\Delta_i}}
\end{equation}
and $C_{ijk}$ are structure constants of three-point correlation functions
\begin{equation}\nn
\left\langle{\cal O}_i(x_1){\cal O}_j(x_2){\cal
O}_k(x_3)\right\rangle=
\frac{C_{ijk}}{|x_1-x_2|^{\Delta_1+\Delta_2-\Delta_3}
|x_1-x_3|^{\Delta_1+\Delta_3-\Delta_2}|x_2-x_3|^{\Delta_2+\Delta_3-\Delta_1}}.
\end{equation}

The HHL three-point functions of two heavy operators and a light
operator can be approximated by a supergravity vertex operator
evaluated at the heavy classical string configuration
\cite{RT2010}: \bea \nn \langle
V_{H}(x_1)V_{H}(x_2)V_{L}(x_3)\rangle=V_L(x_3)_{\rm classical}.
\eea For $\vert x_1\vert=\vert x_2\vert=1$, $x_3=0$, the
correlation function reduces to \bea \nn \langle
V_{H_1}(x_1)V_{H_2}(x_2)V_{L}(0)\rangle=\frac{C_{123}}{\vert
x_1-x_2\vert^{2\Delta_{H}}}. \eea
Then, the structure constants can be given by
\bea \label{nsc}
C_{123}=c_{\Delta}V_L(0)_{\rm classical}, \eea
where $c_{\Delta}$ is the normalization constant of
the corresponding light vertex operator.


\subsection{The string Lagrangian and Virasoro constraints}

String theory moving on certain background can be described by the
Polyakov action \bea\label{PA} &&S^P=-T\int d^2\xi
\sqrt{-\gamma}\gamma^{mn}G_{mn},\h G_{mn}=g_{MN}\p_m X^M \p_n
X^N,\\ \nn &&\p_m=\p/\p \xi^m,\h m,n=(0,1),\h
(\xi^0,\xi^1)=(\tau,\sigma),\h M,N=(0,1,\ldots,9),\eea where $T$
is the string tension. We choose to work in conformal gauge
$\gamma^{mn}=\eta^{mn}=diag (-1,1)$, in which the Lagrangian and
the Virasoro constraints take the form \bea\label{L}
\mathcal{L}_s=T(G_{00}-G_{11}) ,\eea \bea\label{Vir1}
G_{00}+G_{11}=0,\eea \bea\label{Vir2} G_{01}=0.\eea

The background metric $g_{MN}$ for $AdS_4\times CP^3$ is given by
\bea\nn ds^2=g_{MN}dx^M dx^N=R^2\left(ds^2_{AdS^4} +
ds^2_{CP^3}\right), \eea where
$R$ is related to the string tension $T$ and
the 't Hooft coupling constant ($\alpha'=1$) by
\bea\nn TR^2=\sqrt{2\lambda}. \eea
There are also dilaton and
RR-forms, which do not influence the motion of the classical
strings. Further on, we set $R=1$.

The metric of $CP^3$ space can be written as
\bea\label{2p} ds^2_{CP^3}&=& d\theta^2 +
\sin^2 \theta \left(\frac{1}{2}\cos\vartheta_1
d\varphi_1-\frac{1}{2}\cos\vartheta_2
d\varphi_2+d\varphi_3\right)^2 \\ \nn
&+&\cos^2\frac{\theta}{2}\left(d\vartheta_1^2+\sin^2\vartheta_1
d\varphi_1^2\right)
+\sin^2\frac{\theta}{2}\left(d\vartheta_2^2+\sin^2\vartheta_2
d\varphi_2^2\right),\eea where $\theta\in [0,\pi]$,
$\vartheta_1, \vartheta_2 \in [0,\pi]$, $\varphi_1, \varphi_2 \in
[0,2\pi]$, $\varphi_3 \in [0,4\pi]$.
The angular coordinates in (\ref{2p}) can be expressed also by the following
complex coordinates
 \bea\label{z1234} &&z_1= \cos\frac{\theta}{2}
\cos\frac{\vartheta_1}{2} \exp{\left[\frac{i}{2}( \varphi_3
+\varphi_1)\right]}= r_1 \exp{(i\phi_1)},
\\ \nn &&z_2= \sin\frac{\theta}{2}
\cos\frac{\vartheta_2}{2} \exp{\left[-\frac{i}{2}(\varphi_3 -
\varphi_2)\right]} =  r_2 \exp{(i\phi_2)},
\\ \nn &&z_3= \cos\frac{\theta}{2}\sin\frac{\vartheta_1}{2}
\exp{\left[\frac{i}{2}( \varphi_3 -\varphi_1)\right]}=
r_3\exp{(i\phi_3)},
\\ \nn &&z_4= \sin\frac{\theta}{2} \sin\frac{\vartheta_2}{2}
\exp{\left[-\frac{i}{2}(\varphi_3 +\varphi_2)\right]} =
r_4\exp{(i\phi_4)},
\\ \nn &&\sum_{a=1}^{4} r_a^2=1,\h \sum_{a=1}^{4}r_a^2
\p_m\phi_a=0.\eea

\subsection{Giant magnons}

The giant magnon solutions in $CP^3$ can be found by the Neumann-Rosochatius (NR) integrable system
with the following ansatz for the string embedding
\cite{ABR2008} 
\bea\label{NRA} &&t(\tau,\sigma)=\kappa\tau,\h
r_a(\tau,\sigma)=r_a(\xi),\h
\phi_a(\tau,\sigma)=\omega_a\tau+f_a(\xi),\\ \nn
&&\xi=\sigma-v\tau,\h \kappa, \omega_a, v={\rm constants}.\eea

\subsubsection{$CP^1$ giant magnon}

Let us start with the giant magnon living in the $R_t\times CP^1$
subspace. Such subspace can be obtained by setting $\theta=\vartheta_2=\varphi_2=\varphi_3=0$.
What remains is\footnote{This choice of $CP^1=S^2$ subspace corresponds to ``A''-type giant magnon in the ABJM theory
\cite{AhnNep}.
The ``B''-type magnon lives in another $CP^1$ subspace obtained by $\theta=\pi$. Here we consider only A-type.}
\bea\label{CP1r} ds^2=
-dt^2+d\vartheta_1^2+\sin^2\vartheta_1
d\varphi_1^2.\eea
Then (\ref{z1234}) becomes
\bea\label{CP1c} &&z_1= \cos\frac{\vartheta_1}{2}
\exp{\left(\frac{i}{2} \varphi_1\right)}= r_1 \exp{(i\phi_1)},
\\ \nn &&z_2= 0,
\\ \nn &&z_3= \sin\frac{\vartheta_1}{2}
\exp{\left(-\frac{i}{2}\varphi_1\right)}= r_3\exp{(-i\phi_1)},
\\ \nn &&z_4=0.\eea

For this case, the induced metric on the string wordsheet is
\bea\nn &&G_{00}=-(\p_0 t)^2+(\p_0
\vartheta_1)^2+\sin^2 \vartheta_1(\p_0 \varphi_1)^2 ,\\
\nn &&G_{11}=-(\p_1 t)^2+(\p_1
\vartheta_1)^2+\sin^2 \vartheta_1(\p_1 \varphi_1)^2 ,\\
\nn && G_{01}=-\p_0 t\p_1
t+\p_0\vartheta_1\p_1 \vartheta_1+\sin^2
\vartheta_1\p_0 \varphi_1\p_1 \varphi_1.\eea

By using (\ref{NRA}) and (\ref{CP1r}) in (\ref{L}), one can write down the string
Lagrangian in the following form (prime is used for $d/d\xi$)
\bea\label{Lr} \mathcal{L}_s=
-T(1-v^2)\left[(\vartheta_1^{'})^2+
\sin^2\vartheta_1\left(\left(f_1^{'}+\frac{v\omega_1}{1-v^2}\right)^2
-\frac{\omega_1^2}{(1-v^2)^2}\right)\right],\eea
from which the first integral for $f_1$ becomes
\bea\label{fif1} f_1^{'}= \frac{1}{1-v^2}
\left(\frac{C_1}{\sin^2\vartheta_1}-v\omega_1\right)\eea
with an integration constant $C_1$.

The first Virasoro constraint (\ref{Vir1}) along with (\ref{fif1}) becomes
\bea\label{fit1} (\vartheta_1^{'})^2=
\frac{1}{(1-v^2)^2}\left[(1+v^2)\kappa^2-\frac{C_1^2}{\sin^2\vartheta_1}
-\omega_1^2 \sin^2\vartheta_1\right],\eea
while the second constraint (\ref{Vir2}) determines the constant $C_1=\frac{v\kappa^2}{\omega_1}$.
We will further restrict $\omega_1=1$ since we can choose an appropriate unit of $\tau$.
In terms of $\chi\equiv\cos^2\vartheta_1$, (\ref{fit1}) can be written as
\bea\label{rel}
\chi^{'}
=\frac{2}{1-v^2}\sqrt{\chi(\chi_p-\chi)(\chi-\chi_m)},\qquad{\rm with}\quad
\chi_p=1-v^2\kappa^2,\quad \chi_m=
1-\kappa^2.\eea

\subsubsection{$RP^3$ giant magnon}

The $RP^3$ giant magnon lives in the $R_t \times RP^3$ subspace, which can be obtained from (\ref{2p}) by
setting $\vartheta_1=\vartheta_2=\frac{\pi}{2},\ \varphi_3=0$. The
resulting metric is \bea\nn ds^2= -dt^2+ d\theta^2
+\cos^2\frac{\theta}{2} d\varphi_1^2 +\sin^2\frac{\theta}{2}
d\varphi_2^2.\eea 
Correspondingly, the coordinates (\ref{z1234}) reduce to 
\bea\label{comprp3} &&z_1= \frac{1}{\sqrt{2}}\cos\frac{\theta}{2}
 \exp{\left(\frac{i}{2}\varphi_1\right)}= r_1 \exp{(i\phi_1)},
\\ \nn &&z_2= \frac{1}{\sqrt{2}}\sin\frac{\theta}{2}
 \exp{\left(\frac{i}{2}\varphi_2\right)} =  r_2 \exp{(i\phi_2)},
\\ \nn &&z_3= \frac{1}{\sqrt{2}}\cos\frac{\theta}{2}
\exp{\left(-\frac{i}{2}\varphi_1\right)}= r_1\exp{(-i\phi_1)},
\\ \nn &&z_4= \frac{1}{\sqrt{2}}\sin\frac{\theta}{2}
\exp{\left(-\frac{i}{2}\varphi_2\right)} =
r_2\exp{(-i\phi_2)}.\eea

For the case at hand, the metric induced on the string worldsheet
is given by \bea\nn &&G_{00}=-(\p_0 t)^2+(\p_0
\theta)^2+\cos^2\frac{\theta}{2}(\p_0 \varphi_1)^2+ \sin^2\frac{\theta}{2}(\p_0 \varphi_2)^2,\\
\nn &&G_{11}= -(\p_1 t)^2+(\p_1
\theta)^2+\cos^2\frac{\theta}{2}(\p_1 \varphi_1)^2+ \sin^2\frac{\theta}{2}(\p_1 \varphi_2)^2,\\
\nn && G_{01}=-\p_0 t\p_1 t+\p_0\theta\p_1
\theta+\cos^2\frac{\theta}{2}\p_0 \varphi_1\p_1 \varphi_1
+\sin^2\frac{\theta}{2}\p_0 \varphi_2\p_1 \varphi_2.\eea
The string
Lagrangian in this case becomes
\bea\label{Lr2} \mathcal{L}_s=
-T\left[(1-v^2)(\theta^{'})^2-
\cos^2\frac{\theta}{2}\left((vf_1^{'}-\omega_1)^2-f_1^{'}\right)^2
-\sin^2\frac{\theta}{2}\left((vf_2^{'}-\omega_2)^2-f_2^{'}\right)^2\right],\eea
from which the first integrals for $f_1$ and $f_2$ become
\bea\label{fif2}
f_1^{'}= \frac{1}{1-v^2}
\left(\frac{C_1}{\cos^2\frac{\theta}{2}}-v\omega_1\right),\qquad
f_2^{'}= \frac{1}{1-v^2}
\left(\frac{C_2}{\sin^2\frac{\theta}{2}}-v\omega_2\right)
\eea
with integration constants $C_1,\ C_2$.
Since the $RP^3$ giant magnon should be well-defined at $\theta=\pi$,
we impose an extra condition $C_1=0$.

The two Virasoro constraints (\ref{Vir1}) and (\ref{Vir2}) are combined along with (\ref{fif2}) to give a parametric relation
\bea\label{condRP3}
v\kappa^2=C_2{\omega_2},
\eea
and the first integral for $\theta$
\bea\label{fit2} (\theta^{'})^2=
\frac{1}{(1-v^2)^2}\left[(1+v^2)\kappa^2-\omega_1^2-\frac{C_2^2}{\sin^2\frac{\theta}{2}}
+(\omega_1^2-\omega_2^2) \sin^2\frac{\theta}{2}\right].\eea
This time we fix $\omega_2\equiv 1$ and define
\bea\label{chiu}
\chi\equiv\cos^2\frac{\theta}{2},\qquad u\equiv \frac{\omega_1}{\omega_2},
\eea
to rewrite this equation as
\bea\label{solRP3}
\chi'= \frac{ \sqrt{1-u^2}}{1-v^2}
\sqrt{\chi(\chi_p-\chi)(\chi-\chi_m)},\eea where $\chi_p$ and
$\chi_m$ satisfy the equalities: \bea\label{rel1}
\chi_p+\chi_m=\frac{2-(1+v^2)\kappa^2-u^2}{1-u^2},\h\chi_p\chi_m=
\frac{(1-\kappa^2)(1-v^2\kappa^2)}{1-u^2}.\eea

\setcounter{equation}{0}
\section{HHL of dilaton operator with non-zero momentum}
The vertex for the dilaton operator with {\it non-zero}
momentum $j$, originally defined for $AdS_5\times S^5$ in \cite{RT2010},
is modified in the $AdS_4\times CP^3$ case to
\bea\label{Vd} &&V_{ab}^d (j)= \left(\frac{x^m
x_m+z^2}{z}\right)^{-\Delta_d}
(z_az_b)^{j}\left[z^{-2}\left(\p_+x_{m}\p_-x^{m}+\p_+z\p_-z\right)
+\p_+X_{k}\p_-X^{k}\right]\\ \nn &&x^m x_m=-x_0^2+x_i x_i,\h
i=1,2, \eea where we denote the scaling dimension $\Delta_d=4+j$ and $(x^m, z)$
as the Poincare coordinates on $AdS_4$.
The coordinates on $CP^3$ are represented by angular coordinates $X_k$ or equivalently by the complex coordinates $z_a$ 
defined in (\ref{z1234}).
The choice of the indices $(a,b)$ determines the direction of the momentum in the $CP^3$ space.

The $AdS$ part of the giant magnon solution is given by (after Euclidean
rotation, $i \tau=\tau_e$, where $\tau$ is the
worldsheet time) \bea\label{asol} x_{0e}=\tanh(\kappa
\tau_e),\h x_i=0,\h z=\frac{1}{\cosh(\kappa
\tau_e)}.\eea
Replacing (\ref{asol}) into (\ref{Vd}), one finds \bea\label{Vda}
&&V_{ab}^d (j)=
\left(\cosh\kappa\tau_e\right)^{-\Delta_d}
(z_az_b)^{j}\left(\kappa^2
+\p_+X_{k}\p_-X^{k}\right).\eea

\subsection{With the $R_t \times CP^1$ giant magnons}

From (\ref{CP1c}) it is clear that non-vanishing HHL is possible only with the vertex $V_{13}^d (j)$.
Evaluating the Lagrangian on the giant magnon state,
\bea\label{L1}
\p_+X_{k}\p_-X^{k}=\frac{2}{1-v^2}
\left[\chi-1+\frac{1}{2}(1+v^2)\kappa^2\right],\eea
one finds
\bea\label{Vd1} &&V_{13}^d (j)=
\frac{(1-\chi)^{j/2}\left(\kappa^2+\chi-1\right)}{2^{j-1}(1-v^2)
\left(\cosh\kappa\tau_e\right)^{\Delta_d}}.\eea
The normalized structure constant can be obtained by integrating
the vertex over the string worldsheet. \bea\label{sc1}
\mathcal{C}_{13}^{CP^1,d}(j)&=& c_\Delta^d
\int_{-\infty}^{\infty}d\tau_e \int_{-L}^{L}d\sigma V_{13}^d (j)
\\ \nn &=&\frac{c_\Delta^d\sqrt{\pi}}{2^{j-1}(1-v^2)\kappa}
\frac{\Gamma\left(\frac{\Delta_d}{2}
\right)}{\Gamma\left(\frac{\Delta_d+1}{2}\right)}
\int_{\chi_m}^{\chi_p}\frac{d\chi}{\chi^{'}}(1-\chi)^{j/2}\left(\kappa^2+\chi-1\right),
\eea where we replaced the integration over $\sigma$ with
integration over $\chi$ in the following way \bea\nn
\int_{-L}^{L}d\sigma=
2\int_{\chi_m}^{\chi_p}\frac{d\chi}{\chi'},\eea
using Eq.(\ref{rel}) for $\chi'$.
The parameter $L$
is introduced here in order to take into account the giant magnons in
the {\it finite-size} worldsheet volume.

The final expression of the integral in (\ref{sc1}) becomes
\bea\label{sc1f} \mathcal{C}_{13}^{CP^1,d}(j)&=&
c_\Delta^d \frac{\Gamma\left(\frac{\Delta_d}{2}
\right)}{\Gamma\left(\frac{\Delta_d+1}{2}\right)}\frac{\pi^{3/2}}{2^j\kappa}\chi_p^{-1/2}(1-\chi_p)^{j/2}\times
\\ \nn &&\hspace{-3cm}\left[\chi_p
F_1\left(\frac{1}{2};-\frac{1}{2},-\frac{j}{2};1;1-\epsilon,\frac{\chi_p(\epsilon-1)}{1-\chi_p}\right)-(1-\kappa^2)F_1\left(\frac{1}{2};\frac{1}{2},-\frac{j}{2};1;1-\epsilon,\frac{\chi_p(\epsilon-1)}{1-\chi_p}\right)\right] ,
\eea where $F_1(a;b_1,b_2;c;z_1,z_2)$ is a hypergeometric
function of two variables ($Appell\ F_1$) and
\bea\label{defeps}\epsilon=\frac{\chi_m}{\chi_p}.\eea 
We used an integral representation for the $F_1$ in (\ref{sc1f}) \cite{PBM}
 \bea\nn
F_1(a;b_1,b_2;c;z_1,z_2)=\frac{\Gamma(c)}{\Gamma(a)\Gamma(c-a)}
\int_{0}^{1} x^{a-1}(1-x)^{c-a-1}(1-z_1 x)^{-b_1}(1-z_2
x)^{-b_2}dx.\eea
The structure constants are given by the parameters $\kappa,\ v$
which are eventually related to the conserved
angular momentum $J_1$ and the worldsheet momentum $p$ (see
(\ref{rel}), (\ref{defeps})) by
\bea\label{j1} &&J_1= 2T
\sqrt{\frac{1-v^2}{1-v^2 \epsilon}}
\left[\mathbf{K}(1-\epsilon)-\mathbf{E}(1-\epsilon)\right],
\\ \label{p} &&p= 2v \sqrt{\frac{1-v^2 \epsilon}{1-v^2}} \left[\frac{1}{v^2}
\mathbf{\Pi} \left(1-\frac{1}{v^2}\vert
1-\epsilon)\right)-\mathbf{K}(1-\epsilon)\right].\eea Here
$\mathbf{K}$, $\mathbf{E}$ and $\mathbf{\Pi}$ are the complete
elliptic integrals of the first, second, and third kinds, respectively.

\subsubsection{Leading finite-size corrections}

Since $J_1$ and $p$ define the heavy operators of the dual gauge theory,
it is important to express the semiclassical structure constants
$\mathcal{C}_{13}^{CP^1,d}(j)$ in terms of them.
For given $J_1$ and $p$, one can solve numerically (\ref{j1}) and (\ref{p}) to find
corresponding values of $\kappa,\ v$, which can be used to evaluate 
$\mathcal{C}_{13}^{CP^1,d}(j)$ from (\ref{sc1}).

Explicit computations are possible for the case where $J_1$ is large but finite
$J_1\gg T$, equivalently, $\epsilon\ll 1$.
We start by rewriting (\ref{rel}) in the following form
\bea\label{relr} (1+\epsilon)\chi_p = 2-(1+v^2)\kappa^2,
\h \epsilon \chi_p^2=(1-\kappa^2)(1-v^2 \kappa^2).\eea 
Next, we use the small $\epsilon$-expansions 
\bea\label{exp}
&&\chi_p=\chi_{p0}+(\chi_{p1}+\chi_{p2}\log\epsilon) \epsilon,
\\ \nn &&v=v_0+(v_1+v_2\log\epsilon) \epsilon,
\\ \nn &&\kappa^2=1+W_1 \epsilon.\eea
Replacing (\ref{exp}) into (\ref{relr}), one finds relations
\bea\label{s1} \chi_{p0}=1-v_0^2,\h
\chi_{p1}=v_0(v_0-v_0^3-2v_1),\h \chi_{p2}= -2v_0 v_2,\h
W_1=-1+v_0^2 .\eea

The expressions for $v_0, v_1, v_2$ in terms of the worldsheet
momentum $p$ can be found from (\ref{j1}) and (\ref{p})  
\bea\label{vi}
&&v_0= \cos\frac{p}{2}, \\ \nn &&v_1= \frac{1}{4}\sin^2
\frac{p}{2} \cos \frac{p}{2} (1-\log 16), \\ \nn
&&v_2=\frac{1}{4}\sin^2 \frac{p}{2} \cos \frac{p}{2},\eea
along with the expansion parameter in terms of $J_1$ and $p$
\bea\label{epspJ1} \epsilon= 16 \exp
\left(-\frac{J_1}{T \sin \frac{p}{2}}-2\right).\eea

\leftline{\underline{The case of $j=0$}}

Let us begin with the simplest case $j=0$, i.e. dilaton with zero momentum, which is just the 
Lagrangian. 
The $\mathcal{C}_{13}^{CP^1,d}(j)$ in (\ref{sc1f}) simplifies to 
\bea\label{CP10}
\mathcal{C}_{13}^{CP^1,d}(0)= 
\frac{8\sqrt{\pi}c_\Delta^d}{3\kappa\sqrt{\chi_p}}
\left[\chi_p
\mathbf{E}(1-\epsilon)-(1-\kappa^2)\mathbf{K}(1-\epsilon)\right].\eea
Replacing (\ref{vi}), (\ref{epspJ1}) into the $\epsilon$-expantion
of (\ref{CP10}), we obtain (for dilaton with zero momentum
$\Delta_d=4$) \bea\label{CP10r}
\mathcal{C}_{13}^{CP^1,d}(0)\approx \frac{8 c_\Delta^d}{3} \sin
\frac{p}{2}\left[1-4\sin\frac{p}{2}\left(\sin\frac{p}{2}+\frac{J_1}{T}\right)
\exp \left(-\frac{J_1}{T \sin \frac{p}{2}}-2\right)\right].\eea
This is exactly same result as \cite{AB1105}.

\vspace{0.5cm}
\leftline{\underline{The case of $j=$ even integer}}
In this case, the third arguments $-j/2$ of the $Appell\ F_1$ functions in (\ref{sc1f}) are negative integers.
With the help of Mathematica, we have found that these functions can be expressed in terms of the elliptic integrals,
$\mathbf{K}$ and $\mathbf{E}$.
(The $j=0$ result analysed above belongs to this case too.)
We list explicit functional relations for a few simple cases in the Appendix. 

Using small $\epsilon$-expansion of the elliptic integrals, one can find the leading finite-size
corrections of the HHL for any even $j$ in principle.
Here, we present explicit results for $j=2$ ($\Delta_d=6$) as an example:
\bea
\mathcal{C}_{13}^{CP^1,d}(2)&=&\frac{\sqrt{\pi} c_\Delta^d \Gamma\left(\frac{\Delta_d}{2}\right) }
{6 \kappa\sqrt{\chi_p} \Gamma\left(\frac{\Delta_d+1}{2}\right)}
\left[\mathbf{K}(1-\epsilon )
   \left(3 \kappa^2+\chi_p^2 \epsilon -3\right)-\chi_p
\mathbf{E}(1-\epsilon ) (3 \kappa^2+2 (\chi_p+\chi_p
   \epsilon -3))\right]\nn\\
&\approx&
\frac{8 c^{\Delta}_d }{45}
   \sin\frac{p}{2}\ \bigg[2+\cos p\nn\\
   &+&\left.\left(2 \cos p+7 \cos 2
   p-9-\frac{18 J_1}{T}
   \left(\sin\frac{p}{2}+\frac{1}{3} \sin\frac{3 p}{2}\right)\right)e^{-\frac{J1}{T\sin\frac{p}{2}}-2} \right].
   \eea

\leftline{\underline{The case of $j=$ odd integer}}

Since the Appell functions can not written in terms of the elliptic integrals in this case,
we express the $F_1$ in (\ref{sc1f}) using an infinite sum \cite{fw} \bea\label{F1s}
F_1(a;b_1,b_2;c;z_1,z_2)= \sum_{k=0}^{\infty} \frac{(a)_k
(b_2)_k}{(c)_k k!}\ {}_2F_1\left(a+k;b_1;c+k;z_1\right)z_2^k,\quad (a)_k=\frac{\Gamma(a+k)}{\Gamma(a)}.\eea
We can use small $\epsilon$-expansions for the hypergeometric ${}_2F_1$ functions and resum afterward.
The resulting structure constants (\ref{sc1f}) in the leading-order of $\epsilon$ are
\bea\label{sc1e}\mathcal{C}_{13}^{CP^1,d}(j)&\approx& c_\Delta^d
\frac{\sqrt{\pi}\Gamma\left(\frac{\Delta_d}{2}\right)}{ 2^{j-1}\Gamma\left(\frac{\Delta_d+1}{2}\right)}
 \sin\frac{p}{2} \cos^j\frac{p}{2}
\left[{}_2F_1\left(\frac{1}{2};-\frac{j}{2};\frac{3}{2};-\tan^2\frac{p}{2}\right)
+\mathcal{C}\ \epsilon\right]\\
\mathcal{C}&=&\sec^j\frac{p}{2}-\frac{\pi^{3/2}\csc\frac{\pi j}{2}}
{2\Gamma\left(1-\frac{j}{2}\right)\Gamma\left(\frac{1+j}{2}\right)}
-\frac{1}{2}\left(\gamma+\log
4-\frac{J_1}{T}\csc\frac{p}{2}\right)\sec^j\frac{p}{2} \\ \nn
&-&\frac{1}{4}\ {}_2F_1\left(\frac{1}{2};-\frac{j}{2};\frac{3}{2};-\tan^2\frac{p}{2}\right)
\sin\frac{p}{2}\left[(1+3j)\sin\frac{p}{2}+(1+j)\frac{J_1}{T}\right]\\ \nn
&+&\frac{\sqrt{\pi}}{4} \left[2\ {}_2F_1reg^{(0,0,1,0)}
\left(\frac{1}{2};-\frac{j}{2};\frac{1}{2};-\tan^2\frac{p}{2}\right)+{}_2F_1reg^{(0,1,0,0)}
\left(\frac{1}{2};-\frac{j}{2};\frac{1}{2};-\tan^2\frac{p}{2}\right)\right].
\eea
Here $\gamma$ is the Euler's constant and ${}_2F_1\left(a;b;c;z\right)$ the Gauss hypergeometric
function,
\bea\nn
&&{}_2F_1reg(a,b;c;z)= \frac{1}{\Gamma(c)}\ {}_2F_1(a,b;c;z),
\\ \nn &&{}_2F_1reg^{(0,1,0,0)}
\left(a;b;c;z\right)= \frac{\p}{\p b}\ {}_2F_1reg
\left(a;b;c;z\right), \\ \nn &&{}_2F_1reg^{(0,0,1,0)}
\left(a;b;c;z\right)= \frac{\p}{\p c}\ {}_2F_1reg
\left(a;b;c;z\right).\eea

\subsection{With the $R_t \times RP^3$ giant magnons}
From (\ref{comprp3}), all combinations $(a,b)$ in the dilaton vertex (\ref{Vd}) are non-vanishing.
However, we will focus on only $V_{13}^d (j)$ and $V_{24}^d (j)$ for which explicit expressions can be obtained.
Working in the same way as the $CP^1$ case (see (\ref{L1})), one derives  
\bea\nn \p_+X_{k}\p_-X^{k}=
\frac{2(1-u^2)}{1-v^2}\left[\chi-
\frac{1-\frac{1}{2}(1+v^2)\kappa^2}{1-u^2}\right],\eea
which can be used to $V_{13}^d (j)$ and
$V_{24}^d (j)$:
\bea
V_{13}^d (j)&=&\frac{2^{1-j}}{1-v^2}\left(\cosh\kappa\tau_e\right)^{-\Delta_d}
\chi^{j}\left[(1-u^2)\chi-(1-\kappa^2)\right],\label{Vd13}\\
V_{24}^d (j)&=&\frac{2^{1-j}}{1-v^2}\left(\cosh\kappa
\tau_e\right)^{-\Delta_d}
(1-\chi)^{j}\left[(1-u^2)\chi-(1-\kappa^2)\right].
\label{Vd24}
\eea
By integrating $V_{13}^d (j)$ and $V_{24}^d (j)$ over the string worldsheet coordinates,
we derive the corresponding structure constants $\mathcal{C}_{13}^{RP^3,d}$ 
again in terms of the $F_1$ and ${}_2F_1$:
\bea
\mathcal{C}_{13}^{RP^3,d} (j)&=& \frac{c_\Delta^d
\pi^{3/2}\sqrt{1-u^2}}{2^{j-1}\kappa}
\frac{\Gamma\left(\frac{\Delta_d}{2}\right)}{\Gamma\left(\frac{\Delta_d+1}{2}\right)}
\chi_p^{j+1/2}\times\label{C13f}\\
&&\left[\ {}_2F_1\left(\frac{1}{2};-\frac{1}{2}-j;1;1-\epsilon\right)
 -\frac{(1-\kappa^2)}{(1-u^2)\chi_p }\ {}_2F_1\left(\frac{1}{2};\frac{1}{2}-j;1;1-\epsilon\right) \right],
 \nn\\
\mathcal{C}_{24}^{RP^3,d}(j)&=& \frac{c_\Delta^d
\pi^{3/2}\sqrt{1-u^2}}{2^{j-1}\kappa}
\frac{\Gamma\left(\frac{\Delta_d}{2}\right)}{\Gamma\left(\frac{\Delta_d+1}{2}\right)}\ \chi_p^{1/2}
\left(1-\chi_p\right)^{j}\times \label{C24f}\\
&&\hspace{-2cm}
\left[F_1\left(\frac{1}{2};-\frac{1}{2},-j;1;1-\epsilon, \frac{\chi_p(\epsilon-1)}{1-\chi_p}\right)
 -\frac{(1-\kappa^2)}{(1-u^2)\chi_p }\ F_1\left(\frac{1}{2};\frac{1}{2},-j;1;1-\epsilon, 
 \frac{\chi_p(\epsilon-1)}{1-\chi_p}\right) \right],\nn
\eea
where $\chi_p$ is given by (\ref{rel1}).

The parameters $\kappa,\ u,\ v$ in (\ref{C13f}), (\ref{C24f})
are related to the conserved angular momenta $J_1$, $J_2$ and the
worldsheet momentum $p$ along with (\ref{rel1}) by 
\bea\nn
&&J_1= \frac{T \sqrt{\chi_p}}{\sqrt{1-u^2}} \left[\frac{1-v^2
\kappa^2}{\chi_p}\ \mathbf{K}(1-\epsilon)- \mathbf{E}(1-\epsilon)\right],
\\ \nn &&J_2= \frac{T u\sqrt{\chi_p}}{\sqrt{1-u^2}} \mathbf{E}(1-\epsilon),
\\ \nn &&p= \frac{2v}{\sqrt{(1-u^2)\chi_p}} \left[\frac{\kappa^2}{1-\chi_p}
\mathbf{\Pi} \left(-\frac{\chi_p(1-\epsilon)}{1-\chi_p}\vert
1-\epsilon)\right)-\mathbf{K}(1-\epsilon)\right].\eea
For given $J_1,\ J_2,\ p$, one can find corresponding $\kappa,\ u,\ v$, with which
the structure constants can be evaluated. 

A few comments are in order. The structure constants (\ref{C13f})
and (\ref{C24f}) correspond to finite-size dyonic giant magnons
living in the $RP^3$ subspace of $CP^3$. The case of finite-size
giant magnons living in the $RP^2$ subspace can be obtained by
setting $u=0$. For the infinite size case, one can take a limit of $\kappa=1$ and
$\epsilon=0$.
The above results reduce to the zero momentum
dilaton with $j=0$. 
Small $\epsilon$-expansions for $RP^3$ are straightforward for these cases since they are either 
in terms of the hypergeometric ${}_2F_1$ functions or the Appell $F_1$ functions with special
arguments which can be expressed in terms of the elliptic integrals.
We will not present detailed expressions here.

\setcounter{equation}{0}
\section{HHL of primary scalar operators}

The vertex for primary scalar operators is given by \cite{RT2010}
\bea\label{Vpr} &&\hspace{-.5cm}V_{ab}^{pr} (j)= \left(\frac{x^m
x_m+z^2}{z}\right)^{-\Delta_{pr}}
(z_az_b)^{j}\left[z^{-2}\left(\p_+x_{m}\p_-x^{m}-\p_+z\p_-z\right)
-\p_+X_{k}\p_-X^{k}\right]. \eea The scaling dimension is
$\Delta_{pr}=j$.
This reduces for giant magnons (\ref{Vpr}) to
\bea\label{pr}
 \left[\cosh\left(\kappa\tau_e\right)\right]^{-\Delta_{pr}}
(z_a z_b)^{j}\left[\kappa^2\left(\frac{2}{\cosh^2\kappa
\tau_e}-1\right) -\p_+X_{k}\p_-X^{k}\right] .\eea

In the case of the $CP^1$ giant magnons, we consider $V_{13}^{pr}$ with
\bea\label{CP1pr} &&(z_1
z_3)^{j}=\frac{1}{2^j}(1-\chi)^{j/2},
\\ \nn
&&\kappa^2+\p_+X_{k}\p_-X^{k}=
\frac{1}{2(1-v^2)}\left(\kappa^2+\chi-1\right).\eea
By using (\ref{pr}) and (\ref{CP1pr}), we can integrate the vertex
over the string worldsheet to obtain the corresponding structure
constants as follows:
\bea\label{CP1prf}
\mathcal{C}_{13}^{CP^1,pr}(j)&=&
c_{\Delta}^{pr}\frac{\pi^{3/2}}{2^j
\kappa}\frac{\Gamma\left(\frac{\Delta_{pr}}{2}\right)}
{\Gamma\left(\frac{\Delta_{pr}+2}{4}\right)}\chi_p^{-1/2}(1-\chi_p)^{j/2}\times
\\ \nn &&\left\{\left[\frac{\kappa^2\Delta_{pr}}{\Delta_{pr} +1}(1-v^2)+ (1-\kappa^2)\right]
F_1\left(\frac{1}{2};\frac{1}{2},-\frac{j}{2};1;1-\epsilon,\frac{\chi_p(\epsilon-1)}{1-\chi_p}\right)\right.
\\ \nn &&\hspace{5cm}-\left.\chi_p
F_1\left(\frac{1}{2};-\frac{1}{2},-\frac{j}{2};1;1-\epsilon,\frac{\chi_p(\epsilon-1)}{1-\chi_p}\right)\right\}.\eea

The Appell $F_1$ functions in $\mathcal{C}_{13}^{CP^1,pr}(j)$ have the same arguments as in 
$\mathcal{C}_{13}^{CP^1,d}(j)$ in (\ref{sc1f}). 
Therefore, a similar analysis can be done for 
the leading finite-size corrections for
$\mathcal{C}_{13}^{CP^1,pr}(j)$. 
We present here $j=2$ which is the simplest case of even integer ($j=0$ is trivial)
after converting $F_1$ functions into the elliptic functions:
\bea\label{CP1prf2}
&&\mathcal{C}_{13}^{CP^1,pr}(2)\approx \frac{c_{\Delta}^{pr}}{6}
\csc\frac{p}{2} \left\{2\frac{J_1}{T}\sin\frac{p}{2}-(1-\cos
p)\right.
\\ \nn &&+\left.\left[\cos 2p + 4\left(1+\frac{J_1^2}{T^2}\right)\cos p
+4\frac{J_1^2}{T^2}-5+4\frac{J_1}{T}\left(\sin\frac{p}{2}+
\sin\frac{3p}{2}\right)\right]\exp \left(-\frac{J_1}{T \sin
\frac{p}{2}}-2\right)\right\}.\eea

For the HHLs with two $RP^3$ giant magnons, we consider again two primary scalar operators 
$V_{13}^{pr}$ and $V_{24}^{pr}$.
 are given by
 \bea\label{RP3pr13} \mathcal{C}_{13}^{RP^3,pr}(j)&=&
c_{\Delta}^{pr}\frac{\pi^{3/2}\sqrt{1-u^2}}{2^{j-1}\kappa}
\frac{\Gamma\left(\frac{\Delta_{pr}}{2}\right)}
{\Gamma\left(\frac{\Delta_{pr}+2}{4}\right)} \chi_p^{j-1/2}\times
\\ \nn &&\hspace{-2cm}\left\{\left[\frac{\kappa^2\Delta_{pr}}{\Delta_{pr} +1}\frac{1-v^2}{1-u^2}+
\frac{1-\kappa^2}{1-u^2}\right]{}_2F_1\left(\frac{1}{2};\frac{1}{2}-j;1;1-\epsilon\right)
-\chi_p\ {}_2F_1\left(\frac{1}{2};-\frac{1}{2}-j;1;1-\epsilon\right)\right\},\\\label{RP3pr24} \mathcal{C}_{24}^{RP^3,pr}(j)&=&
c_{\Delta}^{pr}\frac{\pi^{3/2}\sqrt{1-u^2}}{2^{j-1}\kappa}
\frac{\Gamma\left(\frac{\Delta_{pr}}{2}\right)}
{\Gamma\left(\frac{\Delta_{pr}+2}{4}\right)}
\chi_p^{-1/2}(1-\chi_p)^{j}\times
\\ \nn &&\hspace{-2cm}\left\{\left[\frac{\kappa^2\Delta_{pr}}{\Delta_{pr} +1}\frac{1-v^2}{1-u^2}+
\frac{1-\kappa^2}{1-u^2}\right]F_1\left(\frac{1}{2};\frac{1}{2},-j;1;1-\epsilon,
\frac{\chi_p(\epsilon-1)}{1-\chi_p}\right)\right.
\\ \nn &&\hspace{5cm}-\left.\chi_p\ F_1\left(\frac{1}{2};-\frac{1}{2},-j;1;1-\epsilon,
\frac{\chi_p(\epsilon-1)}{1-\chi_p}\right)\right\}.\eea

These results are quite similar to those of the dilaton vertex in (\ref{C13f}) and (\ref{C24f}).
As before, the case of $RP^2$ giant magnons can be obtained by taking $u=0$ limit.
One can make small $\epsilon$-expansion by
either using series expansions of ${}_2F_1$ for $V_{13}^{pr}$ or the Appell functions 
which can be expressed by elliptic integrals for $V_{24}^{pr}$.

\setcounter{equation}{0}
\section{HHL of singlet scalar operators with
string levels}

The vertex for singlet scalar operators is given by \cite{RT2010}
\bea\label{VqGM}
V_q= \left[\cosh\left(\kappa\tau_e\right)\right]^{-\Delta_{q}}
\left(\p_+X_{k}\p_-X^{k}\right)^{q},\eea where $q$ is related to
the string level $n$ by $q=n+1$ and
\bea\label{Dq} \Delta_{q}=
2\left(\sqrt{(q-1)\sqrt{\lambda}+1-\frac{1}{2}q(q-1)}+1\right),\eea
with the 't Hooft coupling $\lambda$.

Since the evaluation of $\p_+X_{k}\p_-X^{k}$ for the giant magnons have been given
in previous sections, we just present the results here.

For the $CP^1$ case,
\bea\label{q1} \mathcal{C}^{CP^1}(q)&=&
\frac{c_\Delta^{q}\pi^{3/2}}{2^{q-1}(1-v^2)^{q-1}\kappa}\frac{\Gamma\left(\frac{\Delta_q}{2}\right)}
{\Gamma\left(\frac{\Delta_q+1}{2}\right) } \chi_p^{-1/2} \times
\\ \nn &&\sum_{k=0}^{q}\frac{q!}{k!(q-k)!} \left[\frac{1}{2}(1+v^2)\kappa^2-1\right]^{q-k}
\chi_p^k \
{}_2F_1\left(\frac{1}{2};\frac{1}{2}-k;1;1-\epsilon\right),\eea
For the $RP^3$ case,
\bea\label{q3} \mathcal{C}^{RP^3}(q)&=&
\frac{c_\Delta^{q}\pi^{3/2}(1-v^2)}{\sqrt{(1-u^2)\kappa^2}}\frac{\Gamma\left(\frac{\Delta_q}{2}\right)}
{\Gamma\left(\frac{\Delta_q+1}{2}\right)}\left[\frac{2(1-u^2)}{1-v^2}\right]^q
\chi_p^{-1/2} \times
\\ \nn &&\sum_{k=0}^{q}\frac{q!}{k!(q-k)!} \left[\frac{\frac{1}{2}(1+v^2)\kappa^2-1}{1-u^2}\right]^{q-k}
\chi_p^k \
{}_2F_1\left(\frac{1}{2};\frac{1}{2}-k;1;1-\epsilon\right).\eea
Again, the $RP^2$ giant magnons correspond to $u=0$.

For concrete results, we present small $\epsilon$-expansions for the $CP^1$ giant magnon with a few lower levels as follows:

$q=1$\h (level $n=0$) \bea\label{0} \mathcal{C}^{CP^1}(1)&=&
\frac{c_\Delta^q \sqrt{\pi}\
\Gamma\left(\frac{\Delta_q}{2}\right)}{\Gamma\left(\frac{\Delta_q+1}{2}\right)}
\left\{\sin\frac{p}{2}-\frac{J_1}{2T}-\left[\frac{J_1}{T}\left(5-\cos
p +2\frac{J_1}{T} \csc\frac{p}{2}\right)\right.\right.
\\
\nn &&\left.\left.\hspace{3cm}-2\sin\frac{p}{2} \left(1+\cos
p+\frac{J_1^2}{T^2}\right)\right] \exp \left(-\frac{J_1}{T \sin
\frac{p}{2}}-2\right)\right\},\eea

$q=2$\h (level $n=1$) \bea\label{1} \mathcal{C}^{CP^1}(2)&=&
\frac{c_\Delta^q \sqrt{\pi}\
\Gamma\left(\frac{\Delta_q}{2}\right)}{24 \
\Gamma\left(\frac{\Delta_q+1}{2}\right)} \left\{3 \frac{J_1}{T}-2
\sin\frac{p}{2}+2\left[\frac{J_1}{2T}\left(31+13\cos p
+6\frac{J_1}{2T}\csc\frac{p}{2}\right)\right.\right.
\\ \nn &&\left.\left.\hspace{3cm} -2\left(33 +5\cos p +3\frac{J_1^2}{T^2}\right)
\sin\frac{p}{2}\right] \exp \left(-\frac{J_1}{T \sin
\frac{p}{2}}-2\right)\right\},\eea

$q=3$\h (level $n=2$) \bea\label{3} &&\mathcal{C}^{CP^1}(3)=
\frac{c_\Delta^q \sqrt{\pi}\
\Gamma\left(\frac{\Delta_q}{2}\right)}{480 \
\Gamma\left(\frac{\Delta_q+1}{2}\right)}
\left\{38\sin\frac{p}{2}-15\frac{J_1}{T}\left[12\left(49+57\cos
p-5\frac{J_1^2}{T^2}\cot^2\frac{p}{2}\right)\right.\right.
\\ \nn &&\left.\left.\hspace{5cm}
-2\frac{J_1}{T}(187+97\cos p)\csc\frac{p}{2}\right] \exp
\left(-\frac{J_1}{T \sin \frac{p}{2}}-2\right)\right\}.\eea

\setcounter{equation}{0}
\section{Concluding remarks}
We obtained here some semiclassical normalized structure constants
for strings in $AdS_4\times CP^3$ dual to $\mathcal{N}=6$ super
Chern-Simons-matter theory in three space-time dimensions (ABJM
theory). We considered the cases where the two ``heavy" string
states are finite-size giant magnons living in the $R_t\times
CP^1$, $R_t\times RP^2$ and $R_t\times RP^3$ subspaces and the
``light" states are the dilaton operators
with non-zero momentum, primary scalar operators, and singlet scalar
operators on higher string levels.

Our results can be compared with other computations based on integrability.
One of them is to formulate the HHL structure constants as diagonal form factors of
the light operators with respect to the on-shell particle states corresponding to
the heavy operators \cite{BajJan}.
This approach is particularly useful to understand the {\it finite-size} corrections 
in the HHL in terms of the underlying integrability structure like the 
world-sheet $S$-matrix. 
In this sense, various HHL functions and their finite-size corrections which we have analysed in this paper 
can be used to understand new aspects of integrability in the 
planar limit of $AdS_4/CFT_3$.

\section*{Acknowledgements}
This work is supported in part by the Brain Pool program 171S-1-3-1765
from KOFST and by National Research Foundation of Korea (NRF) grant
(NRF-2016R1D1A1B02007258) (CA) and the NSF grant DFNI T02/6 (PB).
CA also thanks ICTP for a support through Senior Associates program.

\section*{Appendix: Appell $F_1$ functions with special arguments}
We need to evaluate Appell $F_1$ functions $F_1(1/2;\pm 1/2,-j;1;x,y)$.
Using Mathematica, we have found their relations to the elliptic functions $\mathbf{E}, \mathbf{K}$:
\bea
F_1\left(\frac{1}{2};\pm\frac{1}{2},-j;1;x,y\right)=e^{\pm}_j(x,y)\mathbf{E}(x)+k^{\pm}_j(x,y)\mathbf{K}(x),\nn
\eea
where the coefficients $e^{\pm}_j$ and $k^{\pm}_j$ ($j=1,2$) are given by
\bea
e^{+}_1(x,y)&=&\frac{2y}{\pi x},\  e^{-}_1(x,y)=\frac{2(y+3x-6xy)}{3\pi x},\  
k^{+}_1(x,y)=\frac{2(x-y)}{\pi x},\  k^{-}_1(x,y)=\frac{2y(x-1)}{3\pi x},\  \nn\\
e^{+}_2(x,y)&=&\frac{4y(3x-y-xy)}{3\pi x^2},\quad 
e^{-}_2(x,y)=\frac{2(10xy-3xy^2-2y^2+15x^2-20yx^2+8x^2y^2)}{15\pi x^2},\nn\\
k^{+}_2(x,y)&=&\frac{2(3x^2-6x+2y^2+xy^2)}{3\pi x^2},\qquad k^{-}_2(x,y)=\frac{2y(x-1)(y-5x+2xy)}{15\pi x^2}. \nn
\eea


\begin{thebibliography}{99}

\bibitem{AdS/CFT} J. M. Maldacena, ``The large N limit of superconformal
field theories and supergravity,'' Adv. Theor. Math. Phys. {\bf
2}, 231 (1998) [{arXiv:hep-th/9711200}];\\S. S. Gubser, I. R.
Klebanov, and A. M. Polyakov, ``Gauge theory correlators from
non-critical string theory,''
Phys. Lett. {\bf B428}, 105 (1998) [{arXiv:hep-th/9802109}];\\
E. Witten, ``Anti-de Sitter space and holography,'' Adv. Theor.
Math. Phys. {\bf 2}, 253 (1998) [{arXiv:hep-th/9802150}].

\bibitem{review} N. Beisert et al., ``Review of AdS/CFT
Integrability: An Overview,'' Lett. Math. Phys. {\bf 99}, 3 (2012)
[arXiv:hep-th/1012.3982].

\bibitem{BKV} B. Basso, S. Komatsu, and P. Vieira, ``Structure Constants and Integrable Bootstrap in Planar N=4 SYM Theory,''
[arXiv:hep-th/1505.06745].

\bibitem{weakhexagon}B. Eden and A. Sfondrini, ``Three-point functions in N = 4 SYM: the hexagon proposal at three loops,''
JHEP {\bf 1602}, 165 (2016) [arXiv:hep-th/1510.01242];
B. Basso, V. Goncalves, S. Komatsu, and P. Vieira, ``Gluing Hexagons at Three Loops,'' Nucl. Phys. {\bf B907}, 695 (2016) 
 [arXiv:hep-th/1510.01683];
B. Basso, V. Goncalves, and S. Komatsu, ``Structure constants at wrapping order,'' JHEP {\bf 1705}, 124 (2017) 
[arXiv:hep-th/1702.02154].

\bibitem{stronghexagon}Y. Jiang, S. Komatsu, I. Kostov, and D. Serban, ``Clustering and the Three-Point Function,'' [arXiv:hep-th/1604.03575].

\bibitem{hhl} K. Zarembo, ``Holographic three-point functions of
semiclassical states,'', JHEP {\bf 1009}, 030 (2010) [arXiv:hep-th/1008.1059];
M. S. Costa, R. Monteiro, J. E. Santos, and D. Zoakos, ``On three-point correlation functions in the gauge/gravity
duality,'', JHEP {\bf 1011}, 141 (2010) [arXiv:hep-th/1008.1070].

\bibitem{RT2010}
R. Roiban and A. A. Tseytlin, ``On semiclassical computation of
3-point functions of closed string vertex operators in $AdS_5
\times S^5$,'', Phys. Rev. {\bf D82}, 106011 (2010) [arXiv:hep-th/1008.4921].

\bibitem{BajJan}
Z. Bajnok and R. A. Janik, ``Classical limit of diagonal form factors and HHL correlators,'' 
JHEP {\bf 1701}, 063 (2017) [arXiv:hep-th/1607.02830].

\bibitem{ABJM}  O. Aharony, O. Bergman, D. L. Jafferis, and J. Maldacena, ``N = 6 superconformal
Chern - Simons - matter theories, M2-branes and their gravity
duals,'' JHEP {\bf 0810}, 091 (2008) [arXiv:hep-th/0806.1218].

\bibitem{claint} G. Arutyunov and S. Frolov, ``Superstrings on $AdS_4 \times CP^3$ as a Coset Sigma-model,'' 
JHEP {\bf 0809}, 129 (2008) [arXiv:hep-th/0806.4940];
B. Stefanski jr, ``Green-Schwarz action for Type IIA strings on $AdS_4 \times CP^3$,''
Nucl. Phys. {\bf B808}, 80 (2009) [arXiv:hep-th/0806.4948].

\bibitem{GHO}G. Grignani, T. Harmark, and M. Orselli, ``The SU(2) x SU(2) sector in the string dual of N=6 superconformal 
Chern-Simons theory,'' 	Nucl. Phys. {\bf B810}, 115 (2009) [arXiv:hep-th/0806.4959].


\bibitem{ABR2008} C. Ahn, P. Bozhilov, and R.C. Rashkov,
``Neumann-Rosochatius integrable system for strings on $AdS_4
\times CP^3$,'' JHEP {\bf 0809}, 017 (2008) [arXiv:hep-th/0807.3134].

\bibitem{SRyang} S. Ryang, ``Giant magnons and spike solutions with two spins in $AdS_4
\times CP^3$,'' JHEP {\bf 0811}, 084 (2008) [arXiv:hep-th/0809.5106].

\bibitem{AA0811} M. C. Abbott and I. Aniceto,
``Giant Magnons in $AdS_4 \times CP^3$: Embeddings, Charges and a
Hamiltonian,'' JHEP {\bf 0904}, 136 (2009) [arXiv:hep-th/0811.2423].

\bibitem{AR2010} D. Arnaudov and R.C. Rashkov, ``On semiclassical calculation of three-point functions in $AdS_4
\times CP^3$,'' Phys. Rev. {\bf D83}, 066011 (2011) [arXiv:hep-th/1011.4669].

\bibitem{HKY2012} S. Hirano, C. Kristjansen, and D. Young, ``Giant Gravitons on $AdS_4 \times CP^3$ and their
Holographic Three-point Functions,'' JHEP {\bf 1207}, 006 (2012)
[arXiv:hep-th/1205.1959].

\bibitem{BKMO2012} A. Bissi, C. Kristjansen, A. Martirosyan, and M. Orselli, ``On Three-point Functions
in the $AdS_4/CFT_3$ Correspondence,'' JHEP {\bf 1301}, 137 (2013) [arXiv:hep-th/1211.1359].

\bibitem{LGP2012} B.-H. Lee, B. Gwak, and C. Park, ``Correlation functions of the ABJM
model,''  Phys. Rev. {\bf D87}, 086002 (2013) [arXiv:hep-th/1211.5838].

\bibitem{A2013} D. Arnaudov, ``Three-point functions of
semiclassical string states and conserved currents in $AdS_4\times CP^3$,''
Phys. Rev. {\bf D87}, 126004 (2013) [arXiv:hep-th/1302.2119].

\bibitem{AFZ} G. Arutyunov, S. Frolov, and M. Zamaklar, ``Finite-size Effects from Giant
Magnons,'' Nucl. Phys. {\bf B778}, 1 (2007) [arXiv:hep-th/0606126].

\bibitem{HM} D.M. Hofman and J. Maldacena,
``Giant magnons,'' J. Phys. {\bf A39}, 13095 (2006)
[arXiv:hep-th/0604135].



\bibitem{AhnNep} C. Ahn and R. I. Nepomechie, ``N = 6 super Chern-Simons theory S-matrix and all-loop Bethe ansatz equations,''
JHEP {\bf 0809}, 010 (2008) [arXiv:hep-th/0807.1924].

\bibitem{AB0806} C. Ahn and P. Bozhilov, ``Finite-size Effects for Single Spike,''
JHEP {\bf 0807}, 105 (2008) [arXiv:hep-th/0806.1085].

\bibitem{AB1105} C. Ahn and P. Bozhilov, ``Three-point Correlation functions of Giant magnons with finite size,''
Phys. Lett. {\bf B702}, 286 (2011) arXiv:1105.3084 [hep-th].

\bibitem{fw} functions.wolfram.com

\bibitem{PBM} A. P. Prudnikov, Yu. A. Brychkov, and O. I. Marichev, "Integrals and series, vol.3: More special
functions," NY, Gordon and Breach (1990).



\end{thebibliography}
\end{document}